\documentclass[twoside, a4paper, notoc, 11pt ]{JHEP3}
\usepackage{amsfonts}
\usepackage{amssymb}
\usepackage{comment}
\usepackage{epsfig}
\usepackage{graphicx}
\usepackage{amsmath}
\usepackage{amsxtra}

\newcommand{\bea}{\begin{eqnarray}}
\newcommand{\eea}{\end{eqnarray}}
\newcommand{\vo}{{\cal V}}
\newcommand{\be}{\begin{equation}}
\newcommand{\ee}{\end{equation}}
\newcommand{\mc}{\mathcal}

\newcommand{\half}{\frac{1}{2}}
\newcommand{\ti}{\times}

\def\ba{\begin{eqnarray}}
\def\ea{\end{eqnarray}}
\def\nn{\nonumber}
\def\exd{{\rm d}}

\makeatletter
\def\x@arrow{\DOTSB\Relbar}
\def\xlongequalsignfill@{\arrowfill@\x@arrow\Relbar\x@arrow}
\newcommand{\xlongequal}[2]{%
    \ext@arrow 0099\xlongequalsignfill@{#1}{#2}}
\makeatother

\newcommand{\roughly}[1]{\mathrel{\raise.3ex\hbox{$#1$\kern-0.85em
\lower1ex\hbox{$\sim$}}}}

\def\endignore{}
\def\ignore #1\endignore{}
\def\nn{\nonumber}

\def\pref#1{(\ref{#1})}

\def\beq{\begin{equation}}
\def\eeq{\end{equation}}
\def\beqa{\begin{eqnarray}}
\def\eeqa{\end{eqnarray}}

\def\G{\mathcal{G}}

\def\L{\mathcal{L}}
\def\O{\mathcal{O}}
\def\R{\mathcal{R}}
\def\V{\mathcal{V}}

\def\nn{\nonumber}

\def\ssM{{\scriptscriptstyle M}}
\def\ssN{{\scriptscriptstyle N}}

\newcommand{\bmat}{\left(\begin{array}}
\newcommand{\emat}{\end{array}\right)}

\def\endignore{}
\def\ignore #1\endignore{}

\def\-{\hphantom{-}}

\def\s2{\frac{1}{2}}

\def\IF{\relax{\rm I\kern-.18em F}}
\def\II{\relax{\rm I\kern-.18em I}}
\def\IP{\relax{\rm I\kern-.18em P}}
\def\IC{\relax{\rm I\kern-.48em C}}
\def\IR{\relax{\rm I\kern-.18em R}}
\def\IK{\relax{\rm I\kern-.20em K}}
\def\IM{\relax{\rm I\kern-.25em M}}

\def\Dsl{\,\raise.15ex\hbox{/}\mkern-13.5mu D}

\def \one{\relax{\rm 1\kern-.26em I}}

\def\exd{{\rm d}}

\def\KK{{\scriptscriptstyle KK}}
\def\O{\mathcal{O}}

\def\L{\mathcal{L}}

\def\V{\mathcal{V}}

\def\nn{\nonumber}

\def\({\left(}
\def\){\right)}

\title{A Note on the Magnitude of the Flux Superpotential}

\author{Michele Cicoli,$^{1,2,3}$ Joseph P. Conlon,$^{4}$ Anshuman Maharana,$^{5}$ Fernando Quevedo$^{3,6}$\\
$^1$ Dipartimento di Fisica e Astronomia, Universit\`a di Bologna, \\
\quad via Irnerio 46, 40126 Bologna, Italy. \\
$^2$ INFN, Sezione di Bologna, Italy. \\
 $^3$ Abdus Salam ICTP, Strada Costiera 11, Trieste 34014, Italy. \\
$^4$ Department of Theoretical Physics, University of Oxford, 1 Kebble St, Oxford, UK.\\
$^5$ Harish Chandra Research Institute, Chhatnag Road, Jhusi, Allahabad 211 019, India.\\
$^6$ DAMTP, University of Cambridge, Wilberforce Road, Cambridge CB3 0WA, UK.}

\abstract{The magnitude of the flux superpotential $W_{\rm{flux}}$ plays
a crucial r\^ole in determining the scales of IIB string compactifications after moduli stabilisation.
It has been argued that values of $W_{\rm{flux}}\ll 1$ are preferred,
and even required for physical and consistency reasons. This note revisits these arguments.
We establish that the couplings of heavy Kaluza-Klein modes to light states scale with the internal volume as $g\sim M_\KK /M_P\sim \V^{-2/3}\ll 1$
and argue that consistency of the superspace derivative expansion
requires $gF/M^2\sim m_{3/2}/M_\KK \ll 1$, where $F$ is the auxiliary field of the light fields and $M$ the ultraviolet cutoff.
This gives only a mild constraint on the flux superpotential, $W_{\rm{flux}}\ll \V^{1/3}$, which
can be easily satisfied for $\mc{O}(1)$ values of $W_{\rm flux}$. This regime
is also statistically favoured and makes the Bousso-Polchinski mechanism
for the vacuum energy hierarchically more efficient.}

\preprint{DAMTP-2013-48, HRI/ST/1305}

\begin{document}

\tableofcontents

\bigskip

\section{Introduction}

Flux quantisation is a general topological condition on string compactifications. It is the source of a discretuum of vacua after moduli stabilisation which is the basis of the string landscape. The flux superpotential of type IIB string theory
compactified on a Calabi-Yau $X$ is given by \cite{gvw,gkp}:
\be
W_{\rm{flux}}= \int_X \left(F_3-iSH_3\right)\wedge \Omega\,,
\ee
where $F_3$ and $H_3$ are respectively the RR and NSNS three-forms, $S$ is the complex dilaton and $\Omega$ the Calabi-Yau $(3,0)$-form.
Turning on background three-form fluxes fixes $S$ and the complex structure moduli $U$. The value of $W_{\rm{flux}}$ after moduli stabilisation, $W_0$, is determined by a set of integers coming from flux quantisation and it is expected to be of order $\O\left(\sqrt{\frac{\chi}{24}}\right) \simeq \mc{O}(10)$
where $\chi$ is the Euler number of the corresponding F-theory four-fold.
Standard tadpole cancellation conditions set upper bounds on the magnitude of $W_0$ which is usually $W_0\lesssim \mc{O}(100)$.

Even though $W_0$ is determined from quantised fluxes, over the years several arguments have been given that tend to prefer  very small values of $W_0$.\footnote{Notice that in supergravity the superpotential is defined up to a K\"ahler transformation,
and so talking about small or large magnitudes of superpotentials is not a K\"ahler invariant statement.
The K\"ahler invariant quantity with physical meaning is $e^K|W|^2= m_{3/2}^2$. Hence the superpotentials discussed here are in a particular basis.}
These arguments can be grouped as based on either consistency, phenomenology and statistics:
\begin{enumerate}
\item{} {\it Argument from Consistency I}: a small $W_0$ has been required by the following consistency argument.
The use of a derivative expansion in a supersymmetric effective field theory
indicates that there should also be an expansion in powers of $\epsilon\equiv F/M^2$ where $F$ is the auxiliary field of the relevant light fields and $M$ an ultraviolet cutoff. Imposing $\epsilon \ll 1$  implies that the superpotential which is proportional to $F$ should be very small \cite{scrucca,shanta}.

\item{} {\it Argument from Consistency II}: a natural value $W_0\simeq\mc{O}(1-10)$ has been argued to
be incompatible with a four-dimensional effective field theory since it implies background fluxes
with an energy density of order the string scale: $V_{\rm flux}\simeq \mc{O}(M_s^4)$.
This is not true since the important quantity to look at is not the scaling of the flux
potential energy but its vacuum expectation value (VEV). If the dilaton and the complex structure moduli are
fixed supersymmetrically, then this VEV is vanishing at leading order, even if it would formally scale as $M_s^4$.
In order to trust the four-dimensional effective field theory, one has to check that the effects
used to fix the K\"ahler moduli, develop a potential whose VEV satisfies $\langle V\rangle \ll M_{\KK}^4$.

\item{} {\it Argument from Phenomenology I}: a small $W_0$ has been argued to be necessary also for a viable phenomenology.
In the original efforts to stabilise the K\"ahler moduli $T$, a non-perturbative term $W_{\rm np}$ was added to $W_{\rm{flux}}$ \cite{KKLT}. In order to stabilise the $T$-moduli at values large enough to trust the effective field theory,
$W_{\rm np}$ has to be of the same order as $W_{\rm flux}$, requiring the latter to be `fine tuned'
to values as small as $10^{-10}$ in string units. Even though $W_{\rm flux}$ is determined from a combination of integers,
small values of $W_{\rm flux}$ are allowed in the multi-dimensional space of integer fluxes.

\item{} {\it Argument from Phenomenology II}: the string scale $M_s$ is set by the Planck scale $M_P$ and the internal volume $\vo$,
$M_s\simeq M_P/\vo^{1/2}$, whereas the gravitino mass depends also on $W_0$: $m_{3/2}\simeq W_0 M_P/\vo$.
Therefore the standard phenomenological preference for $M_s\simeq M_{\rm GUT}\simeq 10^{16}$ GeV from unification
and $m_{3/2} \simeq M_{\rm soft}\simeq \mc{O}(1)$ TeV in order to address the hierarchy problem, requires
$\vo\simeq \mc{O}(10^4)$ and $W_0\simeq \mc{O}(10^{-11})$.

\item{} {\it Argument from Statistics}: a small $W_0$ has also been argued to be preferred on statistical grounds.
In the original treatments \cite{dd} the magnitude of $W_0$ was argued to be uniformly distributed. More recently, arguments have been given that the statistical distribution of $W_0$ can peak at zero \cite{henry}, indicating some preference for a hierarchically small value of $W_0$. Similarly, recent statements have been made arguing that a small cosmological constant requires a small $W_0$ \cite{dine,banks}.
\end{enumerate}

In this note we revisit these arguments and argue that actually the natural values for the flux superpotential are the largest possible allowed by tadpole constraints, that is $W_0\simeq \mc{O}(10)$. As is well known, the requirement in the original KKLT scenario that
the flux superpotential be the same order as the non-perturbative superpotential no longer holds in the LARGE Volume Scenario (LVS) \cite{LVS,CQS}
where there is also no need to tune $W_0$ to obtain physically interesting scales. More importantly, we will argue that the consistency argument on the derivative expansion, while apparently severe, in fact only imposes the mild constraint $W_0\ll \vo^{1/3}$ which is easily satisfied for generic values of $W_0$ for large enough volume. Finally we also explore the statistics of $W_0$ for uniform distribution and its impact on the Bousso-Polchinski \cite{bp} argument for the cosmological constant and find that large $W_0$ is statistically preferred and in the LVS it substantially improves the tuning needed in the Bousso-Polchinski mechanism.

\section{Arguments from Consistency}

It has been argued that since the supersymmetry multiplet for a chiral superfield includes the auxiliary fields $F$, the standard derivative expansion in an effective field theory will also incorporate an expansion in powers of $F$. Concretely, if heavy fields of mass $M$ have been integrated out, then the effective theory will naively contain an expansion in powers of $F/M^2$ \cite{scrucca,shanta} (here $F$ is the normalised magnitude of the F-term:
$F\equiv \sqrt{K_{T\bar{T}}F^T F^{\bar{T}}}$).

In particular consider string flux compactifications where $\vert F^T \vert \simeq M_P^2 W_0/\vo$. If
the heavy mass is set to be the ten-dimensional Kaluza-Klein scale $M_{\KK} \simeq M_P /\vo^{2/3}$,
then imposing $\vert F \vert /M_{\KK}^2 \ll 1$ would imply:
\be
\frac{\vert F \vert}{M_\KK^2} \simeq W_0 {\cal V}^{1/3} \ll 1.
\ee
As the volume of the compact space has to be large for the effective field theory to be valid, if correct, this condition would set a
strong constraint on the value of $W_0$, implying $W_0 \ll 1$.

We will review this argument here and readdress the original argument for the identification of the right expansion parameter.
As the analysis is made after integrating out the heavy fields, the key-point is that
the physical implications must depend on the strength of the coupling between heavy and light fields.

In order to get some intuition let us discuss this issue in global supersymmetry. We require a case
in which supersymmetry is broken and the F-terms are not trivially small.
We therefore revisit the simplest O'Raifertaigh model.

\subsection{O'Raifertaigh model revisited}

Recall that this model contains three chiral superfields $L,H_1,H_2$ with canonical kinetic terms.
The superpotential looks like:
\be
W = g L\left(H_1^2-m^2\right) + M H_1 H_2,\,
\ee
with $g$ a dimensionless coupling and mass parameters $m,M$ with $m\ll M$.

In the component action the couplings of light ($L$) and heavy ($H_1,H_2$) states can be read explicitly from the scalar potential:
\be
V=g^2 |H_1^2-m^2|^2 + |2g L H_1 +MZ|^2 + M^2 |H_1|^2,
\ee
from which we can easily extract the light-heavy couplings: the cubic coupling $g M L H_1 H_2$ of strength $g M$ and
the quartic coupling $g^2 L^2 H_1^2$ of strength $g^2$.

Supersymmetry is broken since the equations $F_L=F_{H_1}=F_{H_2}=0$ cannot be simultaneously satisfied.
In a simple vacuum with $H_1=H_2=0$ and $F_{H_1}=F_{H_2}=0$ supersymmetry is broken by $F_L=-gm^2$. The splitting of the multiplets is such that for the $H_1$ multiplet the two scalar components acquire masses of order:
\be
M_{\pm} = M \pm \Delta M, \qquad \Delta M=\frac{g^2 m^2}{M}= \frac{g |F|}{M}\,.
\ee
with the fermionic component of mass $M$. Therefore the small expansion parameter is:
\be
\epsilon= \frac{\Delta M}{M}= \frac{g|F|}{M^2}\,.
\ee
If the coupling $g$ were of order one, this would correspond to the naive expansion parameter quoted in the literature $F/M^2$ \cite{scrucca,shanta}.
However in cases of parametrically weak coupling $g$ between light and heavy sectors, this factor can play an important r\^ole.
This will in fact be realised in string compactifications, where the heavy states are Kaluza-Klein modes and their couplings to the light sector will be suppressed by volume factors, as in gravitational interactions.

\subsection{The Kaluza-Klein case}

Let us now consider the case of interest in which the heavy states are Kaluza-Klein (KK) modes of mass
$M_\KK \sim M_s/\vo^{1/6} \sim M_P / \vo^{2/3}$. We will show from three different approaches that the coupling of heavy KK modes to light states is of order $g\sim M_\KK/M_P\sim 1/\vo^{2/3}\ll 1$.

\subsubsection{Naive KK approach}

As $M_{\KK} \sim M_P / \vo^{2/3}$, the Lagrangian for the heavy Kaluza-Klein modes $H$ is:
\be
\mc{L} = - \half \,\partial_{\mu} H \partial^{\mu} H - \half \left( \frac{M_P}{\mc{V}^{2/3}} \right)^2 H^2\,.
\ee
We can then work out the couplings of the heavy KK states with the light volume modulus
by expanding the volume around its VEV as $\vo=\langle \vo \rangle +\delta \vo$.
Writing the canonically normalised light field as $L \sim \delta\vo/\langle \vo \rangle$,
the heavy-light couplings then become:
\be
\delta{\cal L}_{HL}= M_\KK^2 H^2 +\frac{M_\KK}{\langle \vo \rangle ^{2/3}} L H^2 + \frac{1}{\langle \vo \rangle ^{4/3}} L^2 H^2 + \cdots
\ee
From this we can read the (approximate) coupling between the light and heavy fields to be
\be
g\sim \frac{1}{\langle \vo \rangle^{2/3}}\sim \frac{M_\KK}{M_P}\,,
\ee
and we have, similar to the O'Raifertaigh case, $g M_\KK LH^2$ and $g^2 L^2 H^2$ couplings.

\subsubsection{Holomorphy approach}

There is also an argument based on holomorphy of four-dimensional supergravity theories, that can be applied to the higher order couplings as well.
As the light volume modulus ${\rm Re}(T) \sim \vo^{2/3}$ has an axion for its imaginary part, it cannot appear perturbatively in the superpotential.
The superpotential must then read:
\be
W(L,H) =  f_0(U) M_P H^2 + f_1(U) H^3 + f_2(U) \frac{H^4}{M_P} + \ldots
\ee
where $f_i(U)$ are functions of the complex structure moduli. The fact that the mass of $H$ scales as
$M_{\KK} \sim M_P/\vo^{2/3}$ then enforces the unnormalised kinetic terms to look like:
\bea
K & = & - 2 \ln \vo \,+ \mc{V}^{4/3} H \bar{H} + \ldots \\
\mc{L}_{\rm{kinetic}} & = & - \frac{3}{(T + \bar{T})^2} \partial_{\mu} T \partial^{\mu} \bar{T} +  \mc{V}^{4/3} \partial_{\mu} H \partial^{\mu} \bar{H}.
\eea
The interactions between the light modulus $L \sim \ln \vo$ and the heavy KK modes $H$ are all then induced by canonical normalisation.
Prior to normalisation we have:
\be
V_{\rm{unnormalised}} = M_P^2 H^2 + M_P H^3 + H^4 + \frac{H^5}{M_P} + \ldots
\ee
Normalisation then gives:
\be
V_{\rm{normalised}} = M_{\KK}^2 H^2 + M_{\KK} \ti \left( \frac{M_{\KK}}{M_P} \right)^2 H^3
+ \left( \frac{M_{\KK}}{M_P} \right)^4 H^4  + \ldots
\ee
As any power of $\mc{V}$ can be written as $e^{-\lambda L/M_P}$ for some $\lambda$, we extract the heavy-light couplings by expanding the
exponential, to obtain:
\be
V_{HL} = M_{\KK} \left( \frac{M_{\KK}}{M_P} \right) L H^2 + \left( \frac{M_{\KK}}{M_P} \right)^2 L^2 H^2
+ \left( \frac{M_{\KK}}{M_P} \right)^3 L H^3 + \ldots
\ee
We therefore see that any heavy-light interaction is suppressed by the dimensionless coupling $g \sim M_\KK/M_P\sim {\cal{V}}^{-2/3}$.

\subsubsection{Explicit dimensional reduction approach}

Next, we  show  how this coupling arises from dimensional reduction of the higher-dimensional theory. Our discussion will be similar to that in \cite{uber}. In particular we show how integrating out the extra dimensions leads to an effective field theory with cut-off scale $M_\KK$ and dimensionless coupling $g\sim M_\KK/M_P\sim\V^{-2/3}$.

We wish to track how the effective field theory in four dimensions can capture the couplings of massive and massless KK modes
and how they depend on the
underlying scales like $M_\KK$ or $M_s$ or equivalently on the overall volume $\V$.
Starting from the Einstein-Hilbert term in ten dimensions we will extract the volume dependence of the four-dimensional couplings. As usual we will split the metric into background and fluctuations: $ g_{\ssM\ssN} = \overline{g}_{\ssM\ssN} + \kappa h_{\ssM \ssN}$.
We also explicitly scale out the local linear size of
the extra dimensions, $e^{u(x)}$,  from the total metric:
\be
 \hat g_{\ssM\ssN} \exd x^\ssM \exd x^\ssN
 = \omega \, e^{-6u} \,g_{\mu\nu} \exd
 x^\mu \exd x^\nu + e^{2u} g_{mn} \exd y^m \exd y^n +
 \hbox{off-diagonal terms} \,.
\ee
The factor $\omega =M_P^2/M_s^2$
numerically converts to four-dimensional Planck units, and the vacuum value
$\langle e^u \rangle \propto M_s L$ provides a dimensionless
measure of the extra-dimensional linear size ( $\langle
e^{6u} \rangle \simeq (M_s L)^6 \equiv \vo$). Recall the four-dimensional Planck
scale is related to the dimensionless volume, $\V = V
M_s^{6}$, by $M_P^2 = V M_s^{8} = \V M_s^2$, and so
$\omega = {M_P^2}/{M_s^2} = {\V}$, or $M_s \simeq
M_P/\vo^{1/2}$.

Using $\sqrt{- \hat g_{(10)}} = \sqrt{-g_{(4)}} \,
\sqrt{\,g_{(6)}} \; \omega^2 e^{-6u}$ and $\int \exd^6 y \propto
M_s^{-6}$, expanding the action in powers of fluctuations and focussing on
the four-dimensional scalar KK modes $\varphi^i$ contained within $h_{mn}$, gives the following
four-dimensional Einstein term and scalar kinetic terms:
\ba
 -\L_{\rm kin}\, &= &\, M_s^{8} \int \exd^6 y \sqrt{- \hat g_{(10)}} \;
 \hat g^{\mu\nu} \hat\R_{\mu\nu}  =  \omega
 \, M_s^2 \sqrt{-g_{(4)}} \;\left(
 g^{\mu\nu} R_{\mu\nu} +
 g^{\mu\nu} \G_{ij}(\varphi) \partial_\mu \varphi^i \partial_\nu \varphi^j + \cdots \right)\nn\\
 &= & M_P^2 \sqrt{-g_{(4)}} \; g^{\mu\nu}\left(
 R_{\mu\nu} +  \G_{ij}(\varphi)
 \partial_\mu \varphi^i \partial_\nu \varphi^j + \cdots  \right)\,.
\ea
On the other hand the contributions to the scalar potential for the
$\varphi^i$ fields scale as follows:
\ba
 &&-\L_{\rm pot} \simeq M_s^{8}
 \int \exd^6 y \sqrt{- \hat g_{(10)}} \;
 \hat g^{mn} \hat\R_{mn} \nn\\
 && \qquad \quad= \omega^2 \,
 M_s^2 e^{-6u} \sqrt{-g_{(4)}} \;
 (e^{-2u} g^{mn}) f_{ij}(\varphi)
 \partial_m \varphi^i \partial_n \varphi^j + \cdots \nn\\
 && \qquad\quad =
 \frac{M_P^4}{\V^{4/3}} \sqrt{-g_{(4)}} \; U(\varphi) =
 M_\KK^2 M_P^2 \sqrt{-g_{(4)}} \; U(\varphi) \,,
\ea
where we have used $e^{8u} = \V^{4/3}$ and $M_\KK^2/M_P^2 =
(M_\KK^2/M_s^2) (M_s^2/M_P^2) \simeq \V^{-\frac 13} \,
\V^{-1} = \V^{-\frac 43}$.

In terms of the canonically normalised fields $\phi^i$ determined by  $\varphi^i \propto \phi^i /M_P$, we find
the following schematic quadratic, cubic and quartic interactions:
\be
M_\KK^2 M_P^2 \; U\left(\frac{\phi}{M_P}\right) \,= M_\KK^2\, \phi^2+ g M_\KK \phi^3 + g^2\, \phi^4 + \frac{g^3}{M_\KK}\, \phi^5 + \cdots\,,
\ee
with:
\be
g = \frac{M_\KK}{M_P}\sim \frac{1}{\vo^{2/3}}\,.
\ee

Therefore each of the couplings can be clearly written in terms of the dimensionless coupling $g$ and the four-dimensional cut-off scale $M_\KK$. Note that $\phi$ here stands for both light ($L$)  and heavy ($H$) states and therefore this potential captures the light-heavy KK couplings needed in the text to show that in the supersymmetric extension the expansion parameter is
$\epsilon=gF/M^2=F/(M_P M_\KK)\sim W_0/\V^{1/3}$.

Note also that the expansion of the kinetic terms is
of the form $(\partial \phi)^2 + \frac{\phi}{M_P} (\partial \phi)^2 +  \left(\frac{\phi}{M_P^2}\right)^2 (\partial \phi)^2+ \cdots $
illustrating that the low-energy
derivative interactions are Planck suppressed in contrast to those in the
scalar potential that have a universal additional suppression by a
factor of $M_\KK^2/M_P^2 = 1/\V^{4/3}$ relative to the generic
Planck size \cite{uber}. We note that this is in agreement with the results obtained from the requirement of holomorphy of the superpotential.

Having identified $g$ from three different approaches we may then continue  to determine the consistency expansion parameter in terms of $g$ and $F$:
\be
\epsilon = \frac{\Delta M}{M} =\frac{m_{3/2}}{M_\KK}= \frac{W_0}{\vo^{1/3}}= \frac{gF}{M_\KK^2}\,,
\label{eps}
\ee
where we have used that in supergravity theories with (almost) vanishing cosmological constant the value of $F$ is of order
$F\sim W/(M_P \vo)$.\footnote{In no-scale models $F= m_{3/2}M_P$ exactly at leading order.} Imposing $\epsilon\ll 1$ implies:
\be
\frac{W_0}{\vo^{1/3}} \ll 1\,,
\ee
which is easily satisfied for $W_0 \sim \O(1-10)$ and large volume.\footnote{This is consistent with the independent discussion of the validity of the effective field theory in LVS made in section 4.5 of \cite{CQS}.}

The last equality in (\ref{eps}) can be seen as a consistency check that the identification of $g$ and $\epsilon$ are correct since it has the same functional form as in the O'Raifertaigh case. The fact that the expansion parameter is $gF/M^2$ instead of $F/M^2$ makes a big difference since $g\sim\V^{-2/3}$.

\subsection{Integrating out dilaton and complex structure moduli}

A crucial ingredient to stabilise the dilaton $S$ and the complex structure moduli $U$ is the turning
on of three-form background fluxes $G_3$ which carry an energy density of order:
\be
\rho_{\rm flux}=\alpha'^{-4}\int d^6 y \sqrt{g_6} \,G_3\cdot \bar{G}_3 \sim W_0^2 M_s^4\,.
\ee
A natural value $W_0\simeq\mc{O}(1-10)$, then leads to an energy density of order the string scale
which might look incompatible with a four-dimensional effective field theory.
However this argument is too naive since the important quantity to look at is the VEV of the previous expression.
In the context of the four-dimensional supergravity theory, $\rho_{\rm flux}$ can be rewritten as:
\be
\rho_{\rm flux} = V_{\rm flux} = e^K \sum_{S,U} K^{I\bar{J}} D_I W D_{\bar{J}}\bar{W}
\sim \frac{M_P^4}{\vo^2} \,|D_{S,U} W|^2 \sim W_0^2 M_s^4\,.
\ee
Therefore the effective field theory is under control if the $S$ and $U$ moduli are fixed
supersymmetrically by imposing $D_{S,U} W=0$, which implies $\langle V_{\rm flux}\rangle = 0$.
Once the dilaton and the complex structure moduli have been integrated out, one
has then to make sure that subleading effects needed to fix the K\"ahler moduli
develop a potential whose VEV satisfies the constraint $\langle V \rangle \ll M_\KK^4$.

\section{Arguments from Phenomenology}

In general terms, the K\"ahler moduli are not stabilised at tree-level and therefore quantum corrections play an important r\^ole in their stabilisation. The scalar potential depends on the corrections to the superpotential $W$ and the K\"ahler potential $K$.
In the original KKLT scenario for moduli stabilisation, the volume modulus $T$ was stabilised by a superpotential of the form $W = W_0 + A\, e^{-aT}$.
In this case, $W_0 \ll 1$ is required to ensure $T$ is stabilised at large enough values to trust the supergravity approximation.
However, in LVS models the volume mode $T$ can be stabilised at exponentially large values
even with $W_0 \sim \mc{O}(1)$ by exploiting perturbative corrections to the K\"ahler potential.
Stabilisation in the supergravity regime then does not require a small value for $W_0$.

A related argument concerns the gravitino mass $m_{3/2} = e^{K/2} W$. Low-scale supersymmetry can be obtained for $m_{3/2} \ll M_P$,
which might seem to require $W_0 \ll 1$. However this is only true if $e^{K/2}$ is not particularly small.
Given that $K = - 2 \ln \mc{V}$, $e^{K/2}$ gives a factor of $1/\vo$.
If the dimensionless volume in string units is much greater than one, then the gravitino mass
can be much smaller than the Planck scale with $W_0 \sim {\mc{O}}(1)$. This is realised in the LVS framework where a gravitino mass
$m_{3/2} \sim 10^{-15} M_P$ can be achieved with $W_0 \sim 1$ if the volume is stabilised at $\mc{V} \sim 10^{15}$.

\section{Arguments from Statistics}

Let us now consider the statistical distribution of flux compactifications and concentrate on how it depends on the magnitude of $W_0$.
Away from the tail of the distribution, $W_{0}$ is assumed to be uniformly distributed as
a {\it complex} function \cite{dd}.
In \cite{dd} general arguments were given to justify the uniform distribution
of the K\"ahler invariant combination $e^K|W_0|^2$ before K\"ahler moduli stabilisation. This has been confirmed in recent studies of particular Calabi-Yau flux compactifications where complex structure moduli stabilisation was explicitly computed \cite{jose}. The $\O(100)$ moduli were reduced to only two by the use of discrete symmetries to make the problem tractable. 

On the other hand different conclusions have recently been found for a simplified toy-model of the superpotential for the many moduli case without the use of discrete symmetries \cite{henry}. Furthermore, this exploration indicates the possibility that the distribution for the cosmological constant peaks at zero.
However this toy model does not correspond to Calabi-Yau compactifications. In particular, it  assumes a K\"ahler potential for the dilaton and complex structure model of the form
\be
\label{henry1}
K = - \ln (S + \bar{S}) - \sum_{i=1}^{h_{2,1}} \ln (U_i + \bar{U}_i)
\ee
and a linear superpotential
\be
\label{henry2}
W = c_1 + c_2 S + \sum_{i=1}^{h_{2,1}} (a_i + b_i S) U_i.
\ee
These choices lead to a superpotential VEV that is a product of $h_{2,1}$ factors, and the peaking at zero reflects the possibility that any one of these factors may 
be close to zero. 

However, the structure of equations eq. (\ref{henry1}) and (\ref{henry2}) are not those appropriate to Calabi-Yau compactifications, as the complex structure moduli are not separate
and are coupled to each other.
In order to gain a clear understanding of the distribution of $W_0$, a complete treatment of the many moduli case
for explicit Calabi-Yau compactifications should be performed.

In what follows we assume that $W_0$ is uniformly distributed (in the region of values
suitable for moduli stabilisation) and study implications
for the distribution of the cosmological constant. Let us take the density of states to be $\rho$.
Then the number of states in an area $dA$ of the $({\rm Re}(W_0),{\rm Im}(W_0))$-plane is $dN= \rho \,dA$.
Thus the number of states between $|W_0|$ and $|W_0| + \delta |W_0|$ is:
\be
dN= 2 \pi \rho  |W_0|  \delta |W_0|\,.
\label{dN}
\ee
Therefore for uniformly distributed $W_0$ the number of states is proportional to its magnitude $|W_0|$ and a generic $\O(10)$ superpotential is also preferred statistically over the tuned small $W_0$. Given this, let us compare the distribution of the cosmological constant
in the KKLT and LVS constructions. Both of these scenarios give rise to AdS minima and an uplifting is needed.
The uplifting mechanism can be of several sources from anti D3-branes to D-terms, non-perturbative effects, etc.
(see \cite{lift} for a recent review of all these effects in LVS models).
The r\^ole of the uplifting term is to bring the minimum to a value close to Minkowski.
The tuning of the cosmological constant can then be done by variation of the fluxes as in the Bousso-Polchinski scenario.
We will assume the uplifting has been done and concentrate on the  tuning that is achieved by knowing there is an enormous number of flux vacua for a small range of values of $|W_0|$ say from $0$ to $100$.

For KKLT the vacuum energy $\Lambda$ is of order $|W_0|^2/\V^2$. At the minimum of the scalar potential $\vo \sim \ln |W_0|$
and so the vacuum energy behaves as $|W_0|^2$. Now suppose we want the number of states between $\Lambda$ and $\Lambda + \delta\Lambda$.
$\delta \Lambda$ would be $10^{-120} M_P^4$ to have a good spacing to realise the Bousso-Polchinski scenario for the cosmological constant:
\be
\delta\Lambda\sim \frac{2 |W_0|}{\left(\ln|W_0|\right)^2}\,\delta|W_0|\,.
\ee
The number of states between $\Lambda$ and $\Lambda + \delta\Lambda$ is then almost uniform:
\be
dN\sim \rho \left(\ln|W_0|\right)^2\,\delta\Lambda\,,
\ee
since it depends on $|W_0|$ only logarithmically. This means that even though larger values of $|W_0|$ are statistically preferred, for the cosmological constant tuning there is no further difference between small or larger values of $|W_0|$. The tuning of the cosmological constant is achieved as in the original Bousso-Polchinski proposal by having an enormous value of $\rho$ in a uniform distribution of values of $\Lambda$. The relevant quantity for this tuning is $\delta|W_0|$ and not the magnitude of $W_0$.

However for LVS the situation is different. In this case:
\be
\Lambda  \sim \frac{|W_0|^2}{\vo^3}\qquad\text{with}\qquad  \vo\sim |W_0| \,e^{a/g_s},
\ee
implying that the cosmological constant scales as:
\be
\Lambda  \sim \frac{e^{-3a/g_s}}{|W_0|}\,.
\ee
Let us study the distribution at fixed $g_s$. A small shift $\delta |W_0|$ causes a shift in $\Lambda$
of the order:
\be
\delta \Lambda \sim  |W_0|^{-2} e^{-3a/g_s} \delta |W_0|\,,
\ee
and so, by using (\ref{dN}), the number of states between $\Lambda$ and $\Lambda + \delta\Lambda$ is:
\be
dN \sim \rho |W_0|^3 e^{3a/g_s} \delta \Lambda\,,
\label{nv}
\ee
which is not independent of $|W_0|$ anymore.
Thus from the point of view of the cosmological constant,
LVS models have an increasing density as $|W_0|$ increases. (Furthermore, recently in \cite{decay}\ it was found that the decay from a typical vacuum with value $|W_0|$  to another one with a different value of $|W_0|$ is such that
the change in $|W_0|$, $|\Delta W_0|$ (not to be confused with $\delta|W_0|$ above) is also of order ${\mathcal O} (1)$).

These results can also be rephrased in terms of the number $h^{1,2}$ of complex structure
moduli required for efficient tuning of the cosmological constant. Recall that the
number of flux vacua for a Calabi-Yau is given by:
\be
N_{\rm{vac}} \sim N_{\rm{tad}}^{4(h^{1,2}+1)}\,,
\ee
where $N_{\rm{tad}}$ is the number of different values that each flux quanta is allowed to
take. $N_{\rm{tad}}$ is set by tadpole cancellation and is typically of $\mc{O}(10)$, giving:
\be
\rho \sim \frac{N_{{\rm vac}}}{\pi |W_0|_{\rm{max}}^2} \sim 10^{4h^{1,2}}\qquad\text{for}\qquad |W_0|_{{\rm max}}\sim 100\,.
\ee
Using this in \pref{nv} and demanding $\delta N \gtrsim 1 $ for $\delta \Lambda \sim 10^{-120}$
one arrives at:
\be
\label{tune}
10^{4(h^{1,2} - 30)} |W_0|^3 e^{3a/g_s}\gtrsim 1\,.
\ee
Note that $e^{a/g_s} \sim \vo/|W_0| \sim M_P/m_{3/2}$, and so (\ref{tune}) can be rewritten as:
\be
10^{4(h^{1,2} - 30)} |W_0|^3 \left(\frac{M_P}{m_{3/2}}\right)^3\gtrsim 1\,.
\ee
Parameterising the flux superpotential as $|W_0|\sim 10^x$ and the gravitino mass as $M_P/m_{3/2}\sim 10^y$, the previous relation becomes:
\be
h^{1,2} \gtrsim 30 -\frac{3(x+y)}{4}\,,
\ee
implying that in LVS models for a given $y$, larger values of $x$
are more efficient on the tuning of the cosmological constant since they allow the existence of
a reasonable density of states with the observed value of $\Lambda$ for smaller $h^{1,2}$.

\section{Conclusions}

We have revisited the question regarding the magnitude of the flux superpotential in Calabi-Yau compactifications of IIB string theory.

We found that the derivative expansion of supersymmetric field theories with broken supersymmetry
can have a control parameter which is significantly smaller that $F/M^2$. We have estimated
the expansion parameter by three  independent  methods:
a naive Kaluza-Klein reduction, an argument based on holomorphy and a study of the volume dependence of different terms in detailed dimensional reduction. They all give
$g \simeq \vo^{-2/3}$.
This implies that the consistency condition for the effective field theory to be valid is $gF/M^2 \simeq W_0 \ll \vo^{1/3}$
which is satisfied for large volumes without imposing strong constraints on $W_0$. The fact that the coupling among heavy and light KK states is of order $\vo^{-2/3}$ may have other physical implications which may be worth exploring.

We also concluded that for moduli stabilisation, KKLT requires small values of $W_0$ but this is not needed in the LVS context.
A generic value of $W_0\sim \mc{O}(1-10)$ is also preferred statistically and in LVS it makes the Bousso-Polchinski mechanism more efficient.
What is needed to have a small cosmological constant is a small spacing $\delta|W_0|$ and not small $W_0$.
Also, since the scale of the soft terms depends on the values that both $W_0$ and the volume take, the smallness
of $W_0$ is not a necessary criterion for TeV-scale soft terms.
This in keeping with the fact that it is the gravitino mass, determined by $e^{K/2} |W_0|$,  that is K\"ahler invariant while $W_0$ by itself is not.

\section*{Acknowledgements}

We thank the Newton Institute for hospitality during the 'Branes, Strings and M-theory' workshop in 2012 and acknowledge discussions with Cliff Burgess, Shanta de Alwis, Sven Krippendorf,  David Marsh,  Claudio Scrucca, Gary Shiu, Yoske Sumitomo, Marco Serone, Henry Tye and Roberto Valandro. AM thanks ICTP, Trieste and the University of Bologna for hospitality during the last stages of this project.

\appendix


\begin{thebibliography}{99}

\bibitem{gvw}
S.~Gukov, C.~Vafa and E.~Witten,
  ``CFT's from Calabi-Yau four folds,''
  Nucl.\ Phys.\ B {\bf 584} (2000) 69
   [Erratum-ibid.\ B {\bf 608} (2001) 477]
  [hep-th/9906070].

\bibitem{gkp}
S.~B. Giddings, S.~Kachru, and J.~Polchinski, ``Hierarchies from
fluxes in string compactifications,'' {\em Phys. Rev.} {\bf D66} (2002) 106006,
[arXiv:hep-th/0105097].

\bibitem{scrucca}
L.~Brizi, M.~Gomez-Reino and C.~A.~Scrucca,
  ``Globally and locally supersymmetric effective theories for light fields,''
  Nucl.\ Phys.\ B {\bf 820} (2009) 193
  [arXiv:0904.0370 [hep-th]].

\bibitem{shanta}
S.~P.~de Alwis,
  ``Constraints on LVS Compactifications of IIB String Theory,''
  JHEP {\bf 1205} (2012) 026
  [arXiv:1202.1546 [hep-th]].

\bibitem{KKLT}
S.~Kachru, R.~Kallosh, A.~Linde, and S.~P. Trivedi, ``De Sitter
vacua in string theory,'' {\em Phys. Rev.} {\bf D68} (2003) 046005, [arXiv:hep-th/0301240].


\bibitem{dd}
F.~Denef and M.~R.~Douglas,
  ``Distributions of flux vacua,''
  JHEP {\bf 0405} (2004) 072
  [hep-th/0404116].
  ``Distributions of nonsupersymmetric flux vacua,''
  JHEP {\bf 0503} (2005) 061
  [hep-th/0411183];
F.~Denef, M.~R.~Douglas and B.~Florea,
  ``Building a better racetrack,''
  JHEP {\bf 0406} (2004) 034
  [hep-th/0404257].

\bibitem{henry}
Y.~Sumitomo and S.~-H.~H.~Tye,
  ``A Stringy Mechanism for A Small Cosmological Constant,''
  JCAP {\bf 1208} (2012) 032
  [arXiv:1204.5177 [hep-th]].
Y.~Sumitomo and S.~-H.~H.~Tye,
  ``A Stringy Mechanism for A Small Cosmological Constant - Multi-Moduli Cases -,''
  arXiv:1209.5086 [hep-th].

\bibitem{dine}
M.~Bose and M.~Dine,
  ``Gravity Mediation Retrofitted,''
  arXiv:1209.2488 [hep-ph].

\bibitem{banks}
T.~Banks,
  ``The Top $10^{500}$ Reasons Not to Believe in the Landscape,''
  arXiv:1208.5715 [hep-th].

\bibitem{LVS}
V.~Balasubramanian, P.~Berglund, J.~P. Conlon, and F.~Quevedo,
``Systematics of moduli stabilisation in Calabi-Yau flux compactifications,'' {\em JHEP} {\bf 03} (2005) 007,
[arXiv:hep-th/0502058].

\bibitem{CQS}
J.~P.~Conlon, F.~Quevedo and K.~Suruliz,
  ``Large-volume flux compactifications: Moduli spectrum and D3/D7 soft supersymmetry breaking,''
  JHEP {\bf 0508} (2005) 007
  [hep-th/0505076].

\bibitem{bp}
R.~Bousso and J.~Polchinski,
  ``Quantization of four form fluxes and dynamical neutralization of the cosmological constant,''
  JHEP {\bf 0006} (2000) 006
  [hep-th/0004134].

\bibitem{uber}
  C.~P.~Burgess, A.~Maharana and F.~Quevedo,
  ``Uber-naturalness: unexpectedly light scalars from supersymmetric extra dimensions,''
  JHEP {\bf 1105} (2011) 010
  [arXiv:1005.1199 [hep-th]].

\bibitem{jose}
J.~J.~Blanco-Pillado, M.~Gomez-Reino and K.~Metallinos,
  ``Accidental Inflation in the Landscape,''
  arXiv:1209.0796 [hep-th];
J.~Louis, M.~Rummel, R.~Valandro and A.~Westphal,
  ``Building an explicit de Sitter,''
  JHEP {\bf 1210} (2012) 163
  [arXiv:1208.3208 [hep-th]];
D.~Martinez-Pedrera, D.~Mehta, M.~Rummel and A.~Westphal,
  ``Finding all flux vacua in an explicit example,''
  arXiv:1212.4530 [hep-th].

\bibitem{lift}
  M.~Cicoli, A.~Maharana, F.~Quevedo and C.~P.~Burgess,
  ``De Sitter String Vacua from Dilaton-dependent Non-perturbative Effects,''
  JHEP {\bf 1206} (2012) 011
  [arXiv:1203.1750 [hep-th]].

\bibitem{decay}
S.~de Alwis, R.~K.~Gupta, E.~Hatefi and F.~Quevedo,
  ``Stability, Tunneling and Flux Changing de Sitter Transitions in the Large Volume String Scenario,''
  arXiv:1308.1222 [hep-th].


\end{thebibliography}
\end{document}